\documentclass[twocolumn,prl]{revtex4}

\begin{document}

\title{Scharnhorst effect for a general two-parameter lagrangian density}

\author{F. A. Barone\footnote{e-mail: fbarone@cbpf.br}}
\affiliation{Centro Brasileiro de Pesquisas F\'{\i}sicas Rua Dr.\ Xavier Sigaud, 150, Urca, 22290-180 Rio de Janeiro, RJ, Brazil}

\author{C. Farina\footnote{e-mail: farina@if.ufrj.br}}
\affiliation{Instituto de F{\'\i}sica, UFRJ, Caixa Postal 68528, Rio de Janeiro RJ, 21945-970, Brasil}

\date{\today}

\begin{abstract}
We discuss how the propagation of electromagnetic fields described by a general two-parameter lagrangian, which contains the Euler-Heisenberg effective lagrangian and the Born-Infeld lagrangian as particular cases, is affected by a pair of parallel plates that impose boundary conditions in the quantized field. We consider three differents setups, namely: {\bf (i)} two perfectly conducting plates; {\bf (ii)} two infinitely permeable plates and {\bf (iii)} a pair of plates in which one of them is a perfect conductor and the other has an  infinite magnetic permeability. 
\end{abstract}

\maketitle

%

	In contrast to the classical vacuum, the quantum vacuum is not insensitive to external agents, but it behaves as an active medium, where  virtual processes occur giving rise to real physical phenomena. For instance, the vacuum permitivity and permeability will be affected if we consider the radiation field constrained by two parallel and infinitely conducting plates. As a consequence, the velocity of light is changed. This phenomenom, known as the Scharnhorst effect \cite{ScharnhorstPLB90}, can be explained as follows: since the classical field interacts with the radiation field through the fermionic loop any change in the radiation field modes, as those imposed by the presence of the plates, will have some influence on the classical fields. Assuming that the plates do not impose any boundary condition (BC) to the fermionic field, the Scharnhorst effect turns to be a two-loop effect. Scharnhorst concluded that the  speed of light propaganting perpendicularly to the plates (and inside of them) is  greater than the  speed of light  in unconstrained vacuum. Also, for propagation parallel to the plates, he found that the speed of light is unaltered. Scharnhorst's result was reobtained by G. Barton \cite{BartonPLB90} who used a much  simpler technique, based on the Euler-Heisenberg effective lagrangian. Later, this same technique was used  for a discussion of this effect with plates of different nature as well as  the analogous effect in the context of scalar QED  see ref(s) \cite{Cougo-PintoPLB99-Leipzig98-JPA99}. In 1995 Latorre, Pascual e Tarrach \cite{LatorrePascualTarrachNPB95} noted that the results for the speed of light variations obeyed the so called \lq\lq magic formula\rq\rq
\begin {equation}
\label{formulamagica}
{\bar c}=1-{44\over 135}\alpha^{2}{\rho\over m^{4}}\ ,
\end {equation}
where $\bar c$ is the speed of light averaged over all polarizations and directions, $m$ is the electron mass, $\alpha$ is the fine structure constant and $\rho$ is the corresponding  vacuum energy density. The origin of this formula was elucidated by Gies and Dittrich in 1998 \cite{GiesDittrichPLB98}.

	In this paper we shall discuss the Scharnhorst effect assuming that the electromagnetic fields are described by a two-parameter lagrangian density, written as ${\cal L}={\cal L}_{0}+\Delta{\cal L}$, where ${\cal L}_{0}$ is the usual Maxwell lagrangian density, given by ${\cal L}_{0}=-{\cal F}=(1/2)({\bf E}^{2}-{\bf B}^{2})$, and $\Delta{\cal L}$ is a correction term of the form
\begin {equation}
\label {defDeltaLgeral}
\Delta{\cal L}=4{\sigma}{\cal F}^{2}+\beta{\cal G}^{2} 
={\sigma}({\bf E}^{2}-{\bf B}^{2})^{2}+\beta({\bf E}\cdot{\bf B})^{2}\ .
\end {equation}
In the previous equation $\sigma$ and $\beta$ are real constants (two parameters) and ${\cal F}=(1/2)({\bf B}^{2}-{\bf E}^{2})$ and ${\cal G}^{2}=({\bf E}\cdot{\bf B})^{2}$ are the only gauge invariant Lorentz scalars. Classical corrections to the speed of light introduced by general two-parameter lagrangians had been considered in other papers for situations without boundary conditions \cite{NoveloLorenci1,NoveloLorenci2}.

	In particular, if we take ${\sigma}=(e^{4})/(360\pi^{2}m^{4})$ and $\beta=7{\sigma}=(7e^{4})/(360\pi^{2}m^{4})$ in equation (\ref{defDeltaLgeral}), we reobtain the Euler-Heisenberg lagrangian \cite{Schwinger51} (we brought back the constants $\hbar$ and $c$)
\begin {eqnarray}
\label{defL+DeltaL}
&{\cal L}_{EH}&={\cal L}_0+\Delta{\cal L}\cr
&=&{1\over 2}({\bf E}^{2}-{\bf B}^{2})+{1\over 2E_1^{2}}\biggl[{1\over 4}({\bf E}^{2}-{\bf B}^{2})^{2}
+{7\over 4}({\bf E}\cdot{\bf B})^{2}\biggr]\ ,\cr
&{\,}&
\end {eqnarray}
where $\alpha=e^{2}/4\pi\hbar c\cong 1/137$ is the fine structure constant,
$
E_1=[(45\pi^{3}mc^{2})/(2\alpha^{2}\lambda_{C}^{3})]^{1/2}
$
is a constant with dimensions of electric field and $\lambda_{C}=2\pi\hbar/mc$ is the Compton wavelength of the electron. 

Another situation of special interest is a Born-Infeld like lagrangian density
\begin {equation}
\label {defLagrtipoBI}
{\tilde{\cal L}}=E_{0}^{2}\biggl[1-\biggl(1+2{{\cal F}\over E_{0}^{2}}+\gamma{{\cal G}^{2}\over E_{0}^{4}}\biggr)^{1/2}\biggr]\ ,
\end {equation}
where $\gamma$ is a real dimensionless constant and $E_{0}$ is an unknown constant with dimension of an electric field. For $\gamma=-1$ we have the usual Born-Infeld lagrangian \cite{BornInfeldPRSA34}. The Born-Infeld lagrangian was introduced for the first time in order to describe the dynamics of the classical electromagnetic field, avoiding some kinds of problems presented by the Maxweel theory\cite{BornInfeldPRSA34,MieAP12-13}. Some years latter, it reappeared in a completly different context. It is worth noting that the Born-Infeld action was obtained as a low energy action to the vector modes of opened strings theory \cite{FradkinPLB85,TseytlinNPB87,ACNYNPB87}.

	Expanding (\ref{defLagrtipoBI}) in order $E_{0}^{-2}$ we have
\begin {equation}
\label {defBIexpandido}
{\tilde{\cal L}}\cong{1\over 2}({\bf E}^{2}-{\bf B}^{2})+{1\over 2E_{0}^{2}}\Biggl[{1\over 4}({\bf E}^{2}
-{\bf B}^{2})^{2}-\gamma({\bf E}\cdot{\bf B})^{2}\Biggr]\ ,
\end {equation}
which has the same structure of (\ref{defDeltaLgeral}) with ${\sigma}=1/(8E_{0}^{2})$ and $\beta=-\gamma1/(2E_{0}^{2})$.

	In order to compute the Scharnhorst effect for our general lagrangian density (\ref{defDeltaLgeral}), we shall employ the technique developed by G. Barton \cite{BartonPLB90}. It is based on the fact that a correction $\Delta{\cal L}$ to the Maxwell lagrangian density produces a polarization $\bf P$ and a magnetization $\bf M$ which are given, for $\Delta {\cal L}$ written as in Eq. (\ref {defDeltaLgeral}), by
\begin {eqnarray}
\label {fg1a}
P_{i}&=&{1\over 4\pi}{\partial\over\partial E_{i}}\Delta{\cal L}\cr
&=&{1\over 4\pi}\Bigl[4{\sigma}({\bf E}^{2}-{\bf B}^{2})E_{i}+2\beta({\bf E}\cdot{\bf B})B_{i}\Bigr]\ ,
\\\nonumber
\\
\label {fg1b}
M_{i}&=&{1\over 4\pi}{\partial\over\partial B_{i}}\Delta{\cal L}\cr
&=&{1\over 4\pi}\Bigl[-4{\sigma}({\bf E}^{2}-{\bf B}^{2})B_{i}+2\beta({\bf E}\cdot{\bf B})E_{i}\Bigr]\ .
\end {eqnarray}
The main step in Barton's technique consists in substituting, in expressions (\ref {fg1a}) and (\ref {fg1b}), the electromagnetic fields ${\bf E}$ and ${\bf B}$ by a sum of two fields, a classical field and a quantized one: ${\bf E}\longrightarrow {\bf e}+{\bf E} \ , \ {\bf B}\longrightarrow {\bf b}+{\bf B}$. Here, ${\bf e}$ and ${\bf b}$ are classical fields that describe the propagating wave while ${\bf E}$ and ${\bf B}$ are, from now on, quantum field operators, that will be taken as free fields, except by the fact that their modes are altered by the presence of the material plates. In this approximation, it is correct to take the expansion of the quantum fields in terms of the creation and annihilation operators, as usually done in QED. Substututing these expressions for the fields into (\ref {fg1a}) and (\ref {fg1b}), taking the vacuum expectation value, and maintaining only the relevant linear terms in the classical fields ${\bf e}$ and ${\bf b}$, we obtain
\begin {eqnarray}
\label {zdr2}
\langle P_{i}\rangle_{||}&=&{1\over 4\pi}\Bigl[4{\sigma}\Bigl(\langle{\bf{\bf E}}^{2}
-{\bf B}^{2}\rangle_{||}\delta_{ij}\cr
&+&2\langle{E}_{i}{E}_{j}\rangle_{||}\Bigr)+2\beta\langle{B}_{i}
{B}_{j}\rangle_{||}\Bigr]e_{j}=:\chi_{ij}^{(e)}e_{j}\ ,
\\\nonumber
\\
\label {zdr3}
\langle M_{i}\rangle_{||}&=&{1\over 4\pi}\Bigl[4{\sigma}\Bigl(-\langle{{\bf E}}^{2}-{{\bf B}}^{2}\rangle_{||}\delta_{ij}+2\langle{B}_{i}{B}_{j}\rangle_{||}\Bigr)\\
&+&2\beta\langle{E}_{i}{E}_{j}\rangle_{||}\Bigr]b_{j}=:\chi_{ij}^{(m)}b_{j}\ .
\end {eqnarray}
where the symbol $\langle...\rangle_{||}$ means that we are considering the desired boundary conditions, we used the fact that $\langle{ E}_{i}{ B}_{j}\rangle_{||}=0$ and also defined the electric  and magnetic  vacuum polarizabilities  $\chi_{ij}^{(e)}$ and $\chi_{ij}^{(m)}$.
From the quantities $\chi_{ij}^{(e)}$ and $\chi_{ij}^{(m)}$, we obtain the electric permitivity $\epsilon_{ij}=\delta_{ij}+\Delta\epsilon_{ij}$ and magnetic permeability $\mu_{ij}=\delta_{ij}+\Delta\mu_{ij}$
\begin{eqnarray}
\label{def:Deltavarepsilon,Deltamu}
\!\!&\epsilon_{ij}&\!\!=\delta_{ij}+4\pi\chi_{ij}^{(e)}\cr 
\!\!&=&\!\!\delta_{ij}+
4{\sigma}\Bigl[\langle{{\bf E}}^{2}-{{\bf B}}^{2}\rangle_{||}\delta_{ij}+2\langle{ E}_{i}
{ E}_{j}\rangle_{||}\Bigr]+2\beta\langle{ B}_{i}{ B}_{j}\rangle_{||}
\cr\cr
\!\!&\mu_{ij}&\!\!=\delta_{ij}+4\pi\chi_{ij}^{(m)}\cr
\!\!&=&\!\!\delta_{ij}-
4{\sigma}\Bigl[\langle{{\bf E}}^{2}-{{\bf B}}^{2}\rangle_{||}\delta_{ij}-2\langle{ B}_{i}
{ B}_{j}\rangle_{||}\Bigr]+2\beta\langle{ E}_{i}{ E}_{j}\rangle_{||}
\ .\cr
&{\,}&
\end{eqnarray}

	Corrections in $\epsilon_{ij}$ and $\mu_{ij}$ will give rise to a variation in the refraction index $n=(\epsilon\mu)^{1/2}$, given in first order by $\Delta n=1/2(\Delta\epsilon+\Delta\mu)$. This variation implies a velocity of light $c'$ given by $c'=c/n'=c/(n+\Delta n)\cong(1-\Delta n/n)(c/n)=1-\Delta n$, where we used that the free vacuum refraction index $n$ (without boundary conditions) is equal to unity (in our unity system $c=1$).

	With all previous results we can obtain the influence in the speed of light due to the presence of two infinite parallel plates, assuming the dymanics of the electromagnetic fields  described by the Maxwell lagrangian density plus the two-parameter correction (\ref{defDeltaLgeral}). 

	For convenience, let us assume that the plates are parallel to the ${\cal OXY}$ plane, with one of them located at $z=0$ and the other one at $z=L$. Then, let us consider a wave propagating perpendiculary to the plates, that is, in the $\hat z$ direction, and a wave propagating in a direction parallel to the plates, for instance, in the $\hat x$ direction. In both cases, there are two possible polarizations, and in each case, we must take appropriately the electric and magnetic susceptibilities from equations (\ref{def:Deltavarepsilon,Deltamu}), in order to compute the changes in the refractive index, as follows.

\

\begin{itemize}
\item {\bf Propagation parallel to the plates} (${\bf k}=\pm|{\bf k}|{\hat x}$) 
	\begin {enumerate}
\item Polarization in the $\hat y$ direction:
\begin{equation}
{{\bf e}=e_{2}\,{\hat y}\atop {\bf b}=b_{3}\,{\hat z}}\ \ \ \ \
\Longrightarrow\ \ \ \ \ 
{\Delta\varepsilon=\Delta\varepsilon_{22}\atop\Delta\mu=\Delta\mu_{33}}\ ,
\end{equation}
\begin{equation}
\label {Deltan||-y}
\Delta n=4{\sigma}\Bigl(\langle{E}_{2}{ E}_{2}\rangle_{||}+
\langle{B}_{3}{B}_{3}\rangle_{||}\Bigr)+\beta\Bigl(\langle{B}_{2}{B}_{2}\rangle_{||}+\langle{E}_{3}{E}_{3}\rangle_{||}\Bigr)\ .
\end{equation}
%
\item Polarization in the $\hat z$ direction:
\begin {equation}
{{\bf e}=e_{3}{\hat z}\atop {\bf b}=b_{2}{\hat y}}\ \ \ \ \ \Longrightarrow\ \ \ \ 	
\ 
{\Delta\varepsilon=\Delta\varepsilon_{33}\atop\Delta\mu=\Delta\mu_{22}}\ ,
\end {equation}
\begin {equation}
\label {Deltan||-z}
\Delta n=4{\sigma}\Bigl(\langle{ E}_{3}{ E}_{3}\rangle_{||}+
\langle{ B}_{2}{ B}_{2}\rangle_{||}\Bigr)+\beta\Bigl(\langle{ B}_{3}{ B}_{3}\rangle_{||}+\langle{ E}_{2}
{ E}_{2}\rangle_{||}\Bigr)\ .
\end {equation}
\end {enumerate}
\item	{\bf Propagation perpendicular to the plates} \break (${\bf k}=\pm|{\bf k}|{\hat z}$)
\begin {enumerate}
\item Polarization in the $\hat x$ direction:
\begin {equation}
{{\bf e}=e_{1}{\hat x}\atop {\bf b}=b_{2}{\hat y}}\ \ \ \ \ \Longrightarrow\ \ \ \ 
\
{\Delta\varepsilon=\Delta\varepsilon_{11}\atop\Delta\mu=\Delta\mu_{22}}\ ,
\end {equation}
\begin {equation}
\label {Deltanperp-x}
\Delta n=4{\sigma}\Bigl(\langle{ E}_{1}{ E}_{1}\rangle_{||}+
\langle{ B}_{2}{ B}_{2}\rangle_{||}\Bigr)+\beta\Bigl(\langle{ B}_{1}{ B}_{1}\rangle_{||}+\langle{ E}_{2}
{ E}_{2}\rangle_{||}\Bigr)\ .
\end {equation}
\item Polarization in the $\hat y$ direction:
\begin {equation}
{{\bf e}=e_{2}{\hat y}\atop {\bf b}=b_{1}{\hat x}}\ \ \ \ \ \Longrightarrow\ \ \ \ \
{\Delta\varepsilon=\Delta\varepsilon_{22}\atop\Delta\mu=\Delta\mu_{11}}\ ,
\end {equation}
\begin {equation}
\label {Deltanperp-y}
\Delta n=4{\sigma}\Bigl(\langle{ E}_{2}{ E}_{2}\rangle_{||}+
\langle{ B}_{1}{ B}_{1}\rangle_{||}\Bigr)+\beta\Bigl(\langle{ B}_{2}{ B}_{2}\rangle_{||}+\langle{ E}_{1}
{ E}_{1}\rangle_{||}\Bigr)\ .
\end {equation}
\end {enumerate}
\end {itemize}
From now on we shall restrict ourselves to three distinct boundary conditions:

{\bf (i) Two conducting plates-(CC)}

	This configuration refers to two parallel and perfectly conducting plates. The field correlators, submitted to these boundary conditions, are given by \cite{LutkenRavndalPRA85}
\begin {eqnarray}
\label {2renCORRCC}
\langle 0|E_{1}^{2}(x)|0\rangle&=&\ \ \langle 0|E_{2}^{2}(x)|0\rangle\cr
&=&-\langle 0|B_{3}^{2}(x)|0\rangle={\pi^{2}\over 48L^{4}}\biggl(F(\theta)-{1\over 15}\biggr)\ ,\cr
\langle 0|E_{3}^{2}(x)|0\rangle&=&-\langle 0|B_{1}^{2}(x)|0\rangle\cr
&=&-\langle 0|B_{2}^{2}(x)|0\rangle={\pi^{2}\over 48L^{4}}\biggl(F(\theta)
+{1\over 15}\biggr)\ ,\cr
&{\,}&
\end {eqnarray}
where we defined $F(\theta)=3/\sin^{4}(\theta)-2/\sin^{2}(\theta)$ with $\theta=\pi/(Lz)$. 
Substituting results (\ref{2renCORRCC}) into equations (\ref {Deltan||-y}) and (\ref {Deltan||-z}), we obtain a null variation for the refractive index of a wave propagating parallel to the plates: $\Delta n_{\|}^{CC}=0$, which means that there is no variation at all for the speed of light when the propgation is parallel to the plates. For a propagation perpendicular to the plates, equations (\ref{2renCORRCC}), (\ref {Deltanperp-x}) and (\ref {Deltanperp-y}) give $\Delta n_{\perp}^{CC}=-\pi^{2}(4{\sigma}+\beta)/(2^{3}3^{2}5L^{4})$.

{\bf  (ii) Two infinitely permeable plates-(PP)} 

	The electromagnetic field correlators for this set up are given by \cite{ABFTPRA2003}
\begin {eqnarray}
\label {2renCORRPP}
\langle 0|E_{1}^{2}(x)|0\rangle&=& \langle 0|E_{2}^{2}(x)|0\rangle\cr
&=&-\langle 0|B_{3}^{2}(x)|0\rangle=-{\pi^{2}\over 48L^{4}}\biggl(F(\theta)+{1\over 15}\biggr)\ ,\cr
\langle 0|E_{3}^{2}(x)|0\rangle&=&-\langle 0|B_{1}^{2}(x)|0\rangle\cr
&=&-\langle 0|B_{2}^{2}(x)|0\rangle=-{\pi^{2}\over 48L^{4}}
\biggl(F(\theta)-{1\over 15}\biggr)\ .\cr
&{\,}&
\end {eqnarray}
Substituting results (\ref {2renCORRPP}) into equations (\ref{Deltan||-y}), (\ref{Deltan||-z}), (\ref{Deltanperp-x}) and (\ref{Deltanperp-y}), we obtain the same results as those found for the CC conditions presented above, namely:
\begin {equation}
\label {cvb//(PP)}
\Delta n_{\|}^{PP}=0\;\; ;\; 
\Delta n_{\perp}^{PP}=\Delta n_{\perp}^{CC}=-{\pi^{2}\over 2^{3}3^{2}5}{1\over L^{4}}(4{\sigma}+\beta)\ .
\end {equation}

{\bf (iii) A perfectly conducting plate and an infinitely  permeable one -(CP)} 

This configuration refers to a perfectly conducting plate at $z=0$ and an infinitely permeable one at $z=L$. 
The electromagnetic field correlators for this set up are given by \cite{ABFTPRA2003}
\begin {eqnarray}
\label {2renCORRCP}
\langle 0|E_{1}^{2}(x)|0\rangle&=&\ \ \langle 0|E_{2}^{2}(x)|0\rangle\cr
&=&-\langle 0|B_{3}^{2}(x)|0\rangle={\pi^{2}\over 96L^{4}}\biggl(G(\theta)+{7\over 60}\biggr)\ ,\cr
\langle 0|E_{3}^{2}(x)|0\rangle&=&-\langle 0|B_{1}^{2}(x)|0\rangle\cr
&=&-\langle 0|B_{2}^{2}(x)|0\rangle={\pi^{2}\over 48L^{4}}\biggl(G(\theta)
-{7\over 60}\biggr)\ ,\cr
&{\,}&
\end {eqnarray}
where we defined $G(\theta)=6\cos\theta/\mbox{sen}^{4}\theta -     \cos\theta/\mbox{sen}^{2}\theta$. Substituting  (\ref{2renCORRCP}) into equations (\ref{Deltan||-y}), (\ref {Deltan||-z}), (\ref {Deltanperp-x}) and (\ref {Deltanperp-y}), we obtain
\begin {equation}
\label {cvb//(CP)}
\Delta n_{\|}^{CP}=0\; ;\; 
\Delta n_{\perp}^{CP}=-{7\over 8}\Delta n_{\perp}^{CC}={7\pi^{2}\over 2^{6}3^{2}5}{1\over L^{4}}(4{\sigma}+\beta)\ .
\end {equation}

\noindent
From the results established previously we see that, for all the considered boundary conditions, the speed of light is unaltered for propagation parallel to the plates, whatever the coeficients $\sigma$ and $\beta$ of the two-parameter lagrangian density are.

	For a propagation in an arbitrary direction, making an angle $\Theta$ with the normal vector to the plates ($\hat z$ direction), we have $c^{\,\prime}(\Theta)=1-\Delta n_{\perp}\cos^{2}\Theta$. In order to find a \lq\lq magic formula\rq\rq\ for the lagrangian (\ref{defDeltaLgeral}), we avarage the speed ${c}^{\,\prime}(\Theta)$ over all polarizations and directions, a procedure which leads to
\begin{equation}
{\bar c}^{\,\prime}=1-\Delta n_\perp/3\ ,
\end{equation}
and compute the energy density of the electromagnetic field in the vacuum state, $\rho=(1/2)\langle 0|\Bigl({\bf E}^{2}+{\bf B}^{2}\Bigr)|0\rangle$, for the three considered boundary conditions, which can be done with the aid of correlators (\ref{2renCORRCC}), (\ref{2renCORRPP}) and (\ref{2renCORRCP}):
\begin{equation}
\label{2rhos}
\rho^{CC}=\rho^{PP}=-{\pi^{2}\over 2^{4}3^{2}5}{1\over L^{4}}\ \ ,\ \ \rho^{CP}=-{7\over 8}\rho^{CC}\ .
\end{equation}
Now, with the previous results, we can show that for the considered cases
\begin {equation}
\label {formulamagicageral}
{\bar c}^{\,\prime}=1-{2\over 3}(4{\sigma}+2\beta)\rho\ ,
\end {equation}
which is the \lq\lq magic formula\rq\rq\ for the two-parameter lagrangian density ${\cal L}={\cal L}_0+\Delta{\cal L}$, where $\Delta {\cal L}$ is given by equation (\ref{defDeltaLgeral}).

	Choosing coeficients $\sigma$ and $\beta$ to give the Euler-Heisenberg lagrangian (\ref{defL+DeltaL}), we get for the speed of light perpendicular to the plates:
\begin {eqnarray}
\label {cvbc'perp(CC),(PP)}
({c'})_{\perp}^{CC}&=&({c'})_{\perp}^{PP}=1+{11\pi^{2}\over 2^{2}3^{4}5^{2}}{\alpha^{2}\over(mL)^4}\cr
&=&1+{11\alpha^{2}\over 2^{6}3^{4}5^{2}\pi^{2}}\biggl({\lambda_{C}\over L}\biggr)^{4}>1\ ,
\\ \nonumber \\
\label {cvbc'perp(CP)}
({c'})_{\perp}^{CP}&=&1-{7\over 8}{11\pi^{2}\over 2^{2}3^{4}5^{2}}
{\alpha^{2}\over(mL)^4}\cr
&=&1-{77\alpha^{2}\over 2^{9}3^{4}5^{2}\pi^{2}}\biggl({\lambda_{C}\over L}\biggr)^{4}<1\ ,
\end {eqnarray}
where $\lambda_{C}=2\pi/m$ is the Compton wavelength. These results, which are particular cases of ours, are in perfect agreement with those found in literature \cite{ScharnhorstPLB90,BartonPLB90}.

	Taking $\sigma$ and $\beta$ to give the lagrangian density of the Born-Infeld tipe (\ref{defBIexpandido}), we have
\begin {eqnarray}
\label {c'perpBI(CC),(PP)}
({c'})_{\perp}^{CC}=({c'})_{\perp}^{PP}&=&1+{\pi^{2}\over 2^{4}3^{2}5}(1-\gamma){1\over(E_{0}L^{2})^{2}}\ ,\cr
({c'})_{\perp}^{CP}&=&1-{7\pi^{2}\over 2^{7}3^{2}5}(1-\gamma){1\over(E_{0}L^{2})^{2}}\ .
\end {eqnarray}
The magic formula for this Born-Infeld-like lagrangian density  reads (for the averaged velocity of light)
\begin {equation}\label{c'perpBI(CP)}
{\bar c'}=1-{1\over 3}(1-\gamma){\rho\over E_{0}^{2}}
\end {equation}

For a lagrangian density with the general form (\ref{defDeltaLgeral}) we showed that the speed of light for a propagation parallel to the plates isn't changed. For a propagation perpendicular to the plates, the boundary conditions CC and PP give the same  variation for the speed of light. For the CP case, the variation in the speed of light has an opposite sign compared with the results found for the other two ones. In fact, if we take into account equation (\ref{formulamagicageral}) and the vacuum energy densities for the three considered boundary conditions, this result is expected. For a Born-Infeld like lagrangian density, given by (\ref{defBIexpandido}) with a generic coeficient $\gamma$, we can verify according to  (\ref{c'perpBI(CC),(PP)}) or (\ref{c'perpBI(CP)}), the following results.

If we take a factor $\gamma<1$ we will have
\begin {equation}
\gamma<1\ \ \Longrightarrow\ \ ({c'})_{\perp}^{CC}=({c'})_{\perp}^{PP}>1\ \ \ ,\ \ \ ({c'})_{\perp}^{CP}<1\; .
\end {equation}
In the cases where $\gamma>1$, we obtain
\begin {equation}
\gamma>1\ \ \Longrightarrow\ \ ({c'})_{\perp}^{CC}=({c'})_{\perp}^{PP}<1\ \ \ ,\ \ \ ({c'})_{\perp}^{CP}>1\; .
\end {equation}

	However, a very peculiar result occurs when we take $\gamma=1$: if we take $\gamma=1$ we will not have, in the considered approximation (order $1/E_{0}^{2}$), any change in the speed of light due to the presence of the plates. Then, we have
\begin {equation}
\gamma=1\ \ \Longrightarrow\ \ (c')_{\perp}^{CC}=(c')_{\perp}^{PP}=(c')_{\perp}^{CP}=1\ .
\end {equation}
This is an interesting result because, even if we do not take into account quantum effects, the lagrangian density (\ref{defLagrtipoBI}) with $\gamma=1$ will give a null variation for the speed of light  \cite{NoveloLorenci1,NoveloLorenci2}.

	The fact that we obtained values for the speed of light greater than $c=1$ is not in disagreement with causality, since these are phase velocities, and all results presented here are valid only for low frequencies. In order to elucidate this point, one must consider the propagation of a wave-packet and analyse the speed of the wave front. 
	
	As a last comment, we would like to emphasize that Barton's technique is well suited for   calculating variations in the speed of light due to several other reasons, as for example:  thermal effects, other kinds of boundary conditions, etc. All one has to do is to consider the appropriate electromagnetic field correlators.

\bigskip
\noindent
{\bf Acknowledgments}\hfill\break
F.A. Barone would like to acknowledge FAPERJ for financial support and C. Farina would like to 
thank CNPq for partial financial support. The authors would like to thank D.T. Alves and F.C. Santos for suggestions and C. Farina also took profit of a private conversation with C.Wotzasek.

\end{document}